\documentclass[preprint2]{aastex}

\shorttitle{TRGB Distance to NGC300}
\shortauthors{Rizzi et al.}

\begin{document}
\title{The Araucaria Project: The Distance to NGC 300 from the Red Giant Branch Tip using HST/ACS imaging\footnote{Based on observations made with the NASA/ESA Hubble Space Telescope, obtained at the Space Telescope Science Institute, which is operated by the Association of Universities for Research in Astronomy, Inc., under NASA contract NAS 5-26555. These observations are associated with program \# GO-9492.}} 
\author{Luca Rizzi, Fabio Bresolin, Rolf-Peter Kudritzki}
\affil{Institute for Astronomy, 2680 Woodlawn Drive, Honolulu, HI 96822; rizzi, bresolin, kud@ifa.hawaii.edu}

\author{Wolfgang Gieren}
\affil{Departamento de Fisica, Astronomy Group, Universidad de Concepci\'on, Casilla 160-C, Concepci\'on, Chile; wgieren@astro-udec.cl}
\and
\author{Grzegorz Pietrzy\'nski}
\affil{Departamento de Fisica, Astronomy Group, Universidad de Concepci\'on, Casilla 160-C, Concepci\'on, Chile and Warsaw University Observatory, Al. Ujazdowskie 4, PL-00-478, Warsaw, Poland; pietrzyn@hubble.cfm.udec.cl}

\begin{abstract}
We have used the Advanced Camera for Surveys on board the Hubble Space Telescope to obtain deep photometry of the NGC 300 spiral galaxy in the Sculptor group. The results have been used to derive an accurate distance determination based on the Tip of the Red Giant Branch distance estimator. Both edge-detection and maximum likelihood methods have been applied, to derive a distance modulus
$(m-M)_0=26.30 \pm 0.03 \pm 0.12 $ for edge-detection, and $(m-M)_0=26.36 \pm 0.02  \pm 0.12$ for maximum likelihood. These results are fully consistent with the recent distance estimate derived from near-IR photometry of Cepheids variable stars in the context of the Araucaria project, $(m-M)_0= 26.37 \pm 0.05 \pm 0.03$ \citep{2005ApJ...628..695G}. 
\end{abstract}

\keywords{galaxies: distances and redshifts --- galaxies: individual (NGC300)}

\section{Introduction}

The brightest of five main spiral galaxies that form the Sculptor group, NGC 300 is a fairly typical late-type galaxy \citep{1988ngc..book.....T} at a distance of $\sim 2.1$ Mpc \citep{1992ApJ...396...80F}.

Most of the measurements of the distance to this galaxy are based on the luminosity of its Cepheid variables population. Based on near-infrared H-band observations of two long-period Cepheid variables, \citet{1987ApJ...320...26M} reported a distance modulus  $(m-M)_0=26.35 \pm 0.25$. The distance was slightly revised by \citet{1988PASP..100..949W} who derived a distance modulus  $(m-M)_0=26.4 \pm 0.2$. Additional photometry of the same sample of variables by \citet{1992ApJ...396...80F} resulted in the already quoted distance  $(m-M)_0=26.66 \pm 0.10$, subsequently revised to $(m-M)_0=26.63 \pm 0.06$ in \citet{2004ApJ...608...42S}.

More recently, NGC 300 has been selected as a key target for the Araucaria Project\footnote{http://ifa.hawaii.edu/$\sim$bresolin/Araucaria}. \citet{2002AJ....123..789P} presented an extensive characterization of 117 Cepheid variables, most of which were new discoveries, observed with the 2.2m ESO/MPI telescope at La Silla, Chile. Additional V and I data were obtained by \citet{2004AJ....128.1167G} at Las Campanas and Cerro Tololo. Deep, near-infrared J and K band observations were obtained with ESO VLT using the ISAAC camera, resulting in a final distance modulus  $(m-M)_0=26.37 \pm 0.05 \,{\rm (random)} \pm 0.03 \,{\rm (systematic)}$ \citep{2005ApJ...628..695G}. 

%The superb angular resolution offered by the Hubble Space Telescope has recently open the possibility of determining the distance to NGC 300 using the Tip of the Red Giant Branch (TRGB). Using HST WFPC2 photometry \citet{2004AJ....127.1472B} derived a distance modulus  $(m-M)_0=26.56 \pm 0.07 \pm 0.13$. The same archive HST WFPC2 data were re-analised by \citet{2005A&A...431..127T}, to derive a distance modulus  $(m-M)_0=26.50 \pm 0.15$.

The superb angular resolution offered by the Hubble Space Telescope has recently open the possibility of determining the distance to NGC 300 using the Tip of the Red Giant Branch (TRGB). A set of HST WFPC2 fields were analyzed by \citet{2004ApJ...608...42S}  and \citet{2004AJ....127.1472B}, and more recently by \citet{2005A&A...431..127T}. The derived distance moduli are $(m-M)_0=26.65 \pm 0.09$, 
$(m-M)_0=26.56 \pm 0.07 \pm 0.13$, and $(m-M)_0=26.50 \pm 0.15$, respectively. 

In this paper, we present the first TRGB distance based on deep ACS observations of NGC 300. These data are the deepest ever obtained for this galaxy, and they sample both the inner bulge and the outer disk. The paper is organized as follows: Section \ref{data} presents the data, the reduction techniques we adopted, and the resulting color-magnidute diagrams (CMD). We describe the TRGB method and its application to NGC 300 in Section \ref{distance}. We discuss our results in Section \ref{discussion} and a brief summary is presented in Section \ref{conclusions}.

\section{Observations, Data reduction, and Color-Magnitude Diagrams}
\label{data}

The ACS observations used to derive a new TRGB distance to NGC 300 were obtained during HST Cycle 11, as part of program GO-9492 (PI: Bresolin), from July 2002 to December 2002. The main purpose of these observations was to complement the extensive ground-based CCD photometry of Cepheid variable stars and blue supergiant stars collected in the framework of the Araucaria project. 
Two-orbit HST visits allowed to obtain deep photometry in the F435W, F555W (1080 seconds), and F814W (1440 seconds) filters. A total of six fields were observed. 

Stellar photometry was performed with the DOLPHOT (version 1.0) package, an adaptation of HSTphot 
\citep{2000PASP..112.1383D} to ACS images. Pre-computer Point Spread Functions were adopted, and the final calibrated photometry was then transformed to the standard BVI system using the equations provided by  \citet{2005astro.ph..7614S}.  The transformation from one photometric system to another inevitably introduces additional uncertainties but it seems necessary given that most of the calibrations of the absolute magnitude of the TRGB are in the I band.  For a more extended discussion of the issues related to calibration see \citet{Bresolin:rb}.

\begin{figure}
\plotone{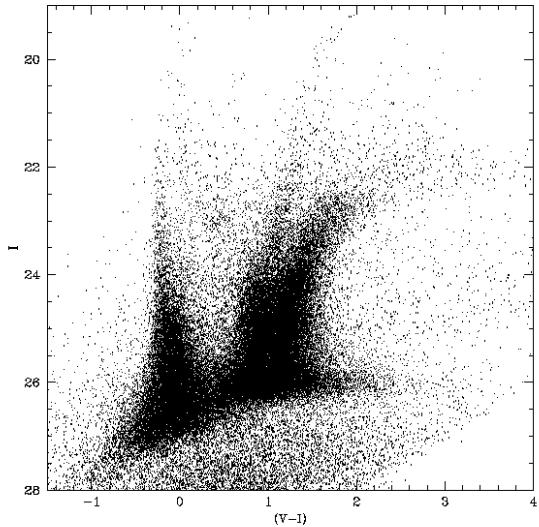}
\caption{(V-I,I) color-magnitude diagram for Field 1 of NGC 300. The Field is situated close to the eastern edge of the galaxy.}
\label{cmd_1.ps}
\end{figure}

\begin{figure}
\plotone{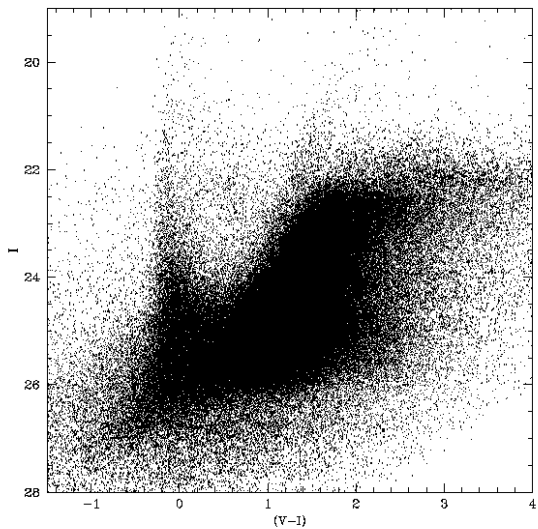}
\caption{(V-I,I) color-magnitude diagram for Field 3 of NGC 300. The Field is situated on top of the nucleus of the galaxy.}
\label{cmd_3.ps}
\end{figure}

As an example of the quality of the results, the final calibrated CMDs are shown in Figures \ref{cmd_1.ps} and \ref{cmd_3.ps} for Fields 1 and 3, respectively. Field 1 is situated close to the eastern outer edge, while Field 3 is centered on the nucleus of the galaxy \citep[see][for a map of the observed Fields]{Bresolin:rb}. All the CMDs show a very well pronounced sequence of blue young stars, reaching down to the lower age limit of isochrone sets \citep[$\sim 60$ Myr,][]{2000A&AS..141..371G}. Blue-loop stars occupy the central region of the diagrams, and a well defined red giant branch (RGB) extends from I $\sim 22$ down to the photometric detection limit, I $\sim 26$. 
A full discussion of the CMD features, along with a reconstruction of the star formation history, will be presented in a forthcoming paper.

\section{The Distance to NGC300}
\label{distance}
\subsection{Detection of the tip}
The distance estimates based on the RGB tip rest on a solid physical basis: low-mass stars reach the end of their ascent along the RGB with a degenerate helium core and they ignite helium burning within a very narrow range of luminosity \citep[][and references therein]{2002PASP..114..375S}.
The potential of the method was revealed in a seminal paper by Lee and collaborators \citep{1993ApJ...417..553L}, along with a first attempt at objectively estimate the position of the tip on the CMD based on a digital edge-detection (ED) filter in the form [-2,0,2], applied to the I band luminosity function. This filter effectively responds to changes in the slope of the luminosity function and displays a peak corresponding to the TRGB. A refined version of this method was presented in \citet{1996ApJ...461..713S}. More recently, a different approach was suggested by \citet{2002AJ....124..213M}. To avoid problems related to binning, this method uses a maximum-likelihood (ML) analysis to get the best fit of a parametric RGB luminosity function to the observed one. Each of these methods has advantages and disadvantages. ED methods are quite sensitive to binning, but they don't require any {\it a priori} assumption on the shape of the RGB luminosity function. ML methods use much more information, because every star of the sample contributes to the probability distribution, but they use a theoretical luminosity function as an input parameter. In this work, we use both approaches and we discuss the different results. 

Whenever color information is available, it is advisable to restrict the analysis of the luminosity function to a suitable region carefully chosen to represent the RGB.
To perform this selection, we took into account the available calibrations of the absolute magnitude of the RGB tip. As discussed in Section \ref{calib}, one of the most reliable calibrations available to date is based on the absolute magnitude of the RGB tip measured on a large sample of stars of the globular cluster $\omega$ Centauri \citep{2001ApJ...556..635B}, at a metallicity of $\rm{[Fe/H]} \sim -1.7$. To be able to apply this calibration to our data, we decided to select our RGB stars using the ridge line of $\omega$ Centauri and selecting stars in a narrow ($\sim$0.1 mag) range on both sides of it. Section \ref{calib} will present a discussion of the implications of this choice.

\begin{figure}
\plotone{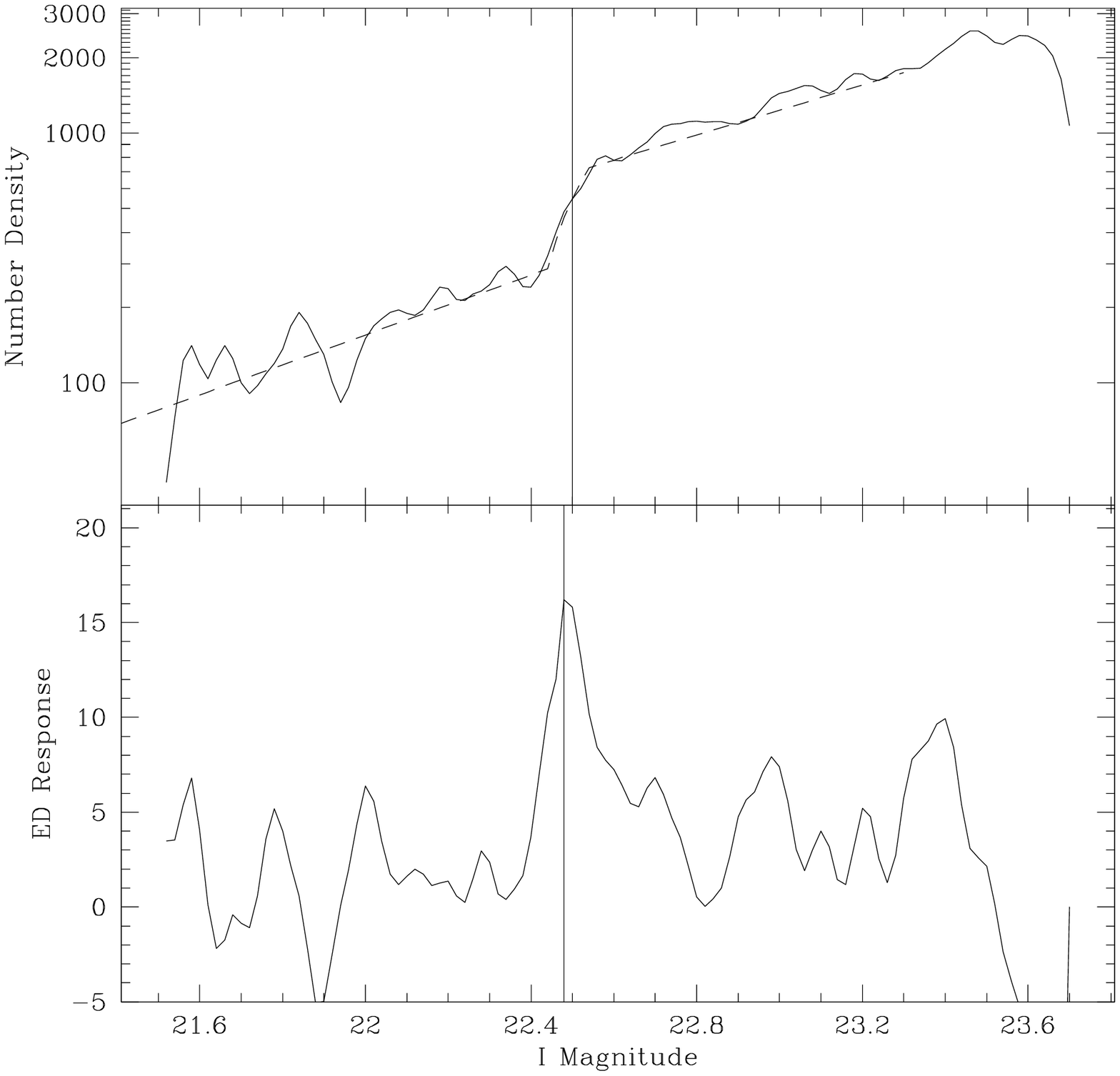}
\caption{Upper panel: Detection of the TRGB using ML method applied to Field 1. Lower panel: Detection of the TRGB using ED method applied to Field 1.}
\label{trgb_1.ps}
\end{figure}

\begin{figure}
\plotone{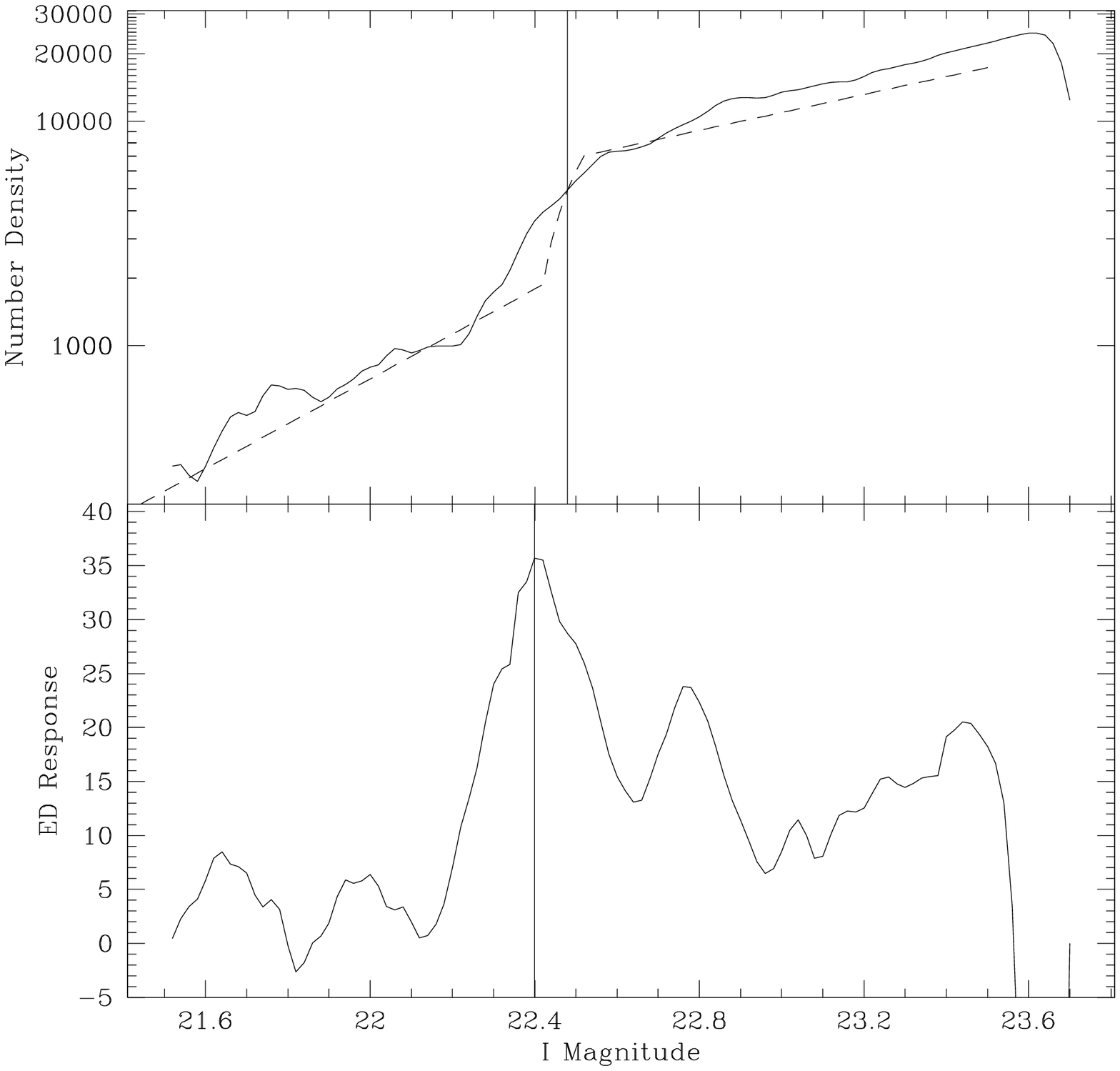}
\caption{Upper panel: Detection of the TRGB using ML method applied to Field 2. Lower panel: Detection of the TRGB using ED method applied to Field 2.}
\label{trgb_2.ps}
\end{figure}

\begin{figure}
\plotone{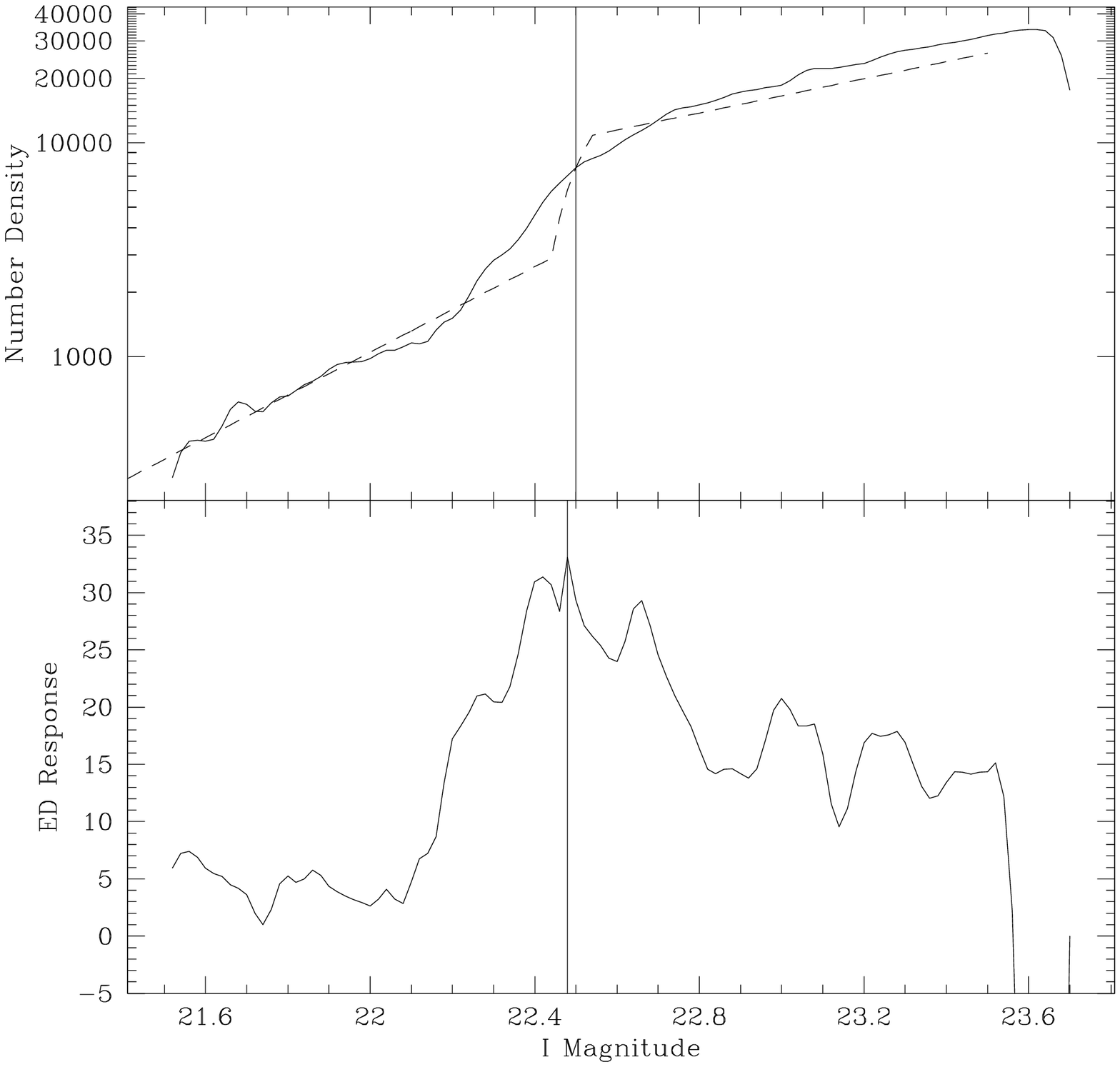}
\caption{Upper panel: Detection of the TRGB using ML method applied to Field 3. Lower panel: Detection of the TRGB using ED method applied to Field 3.}
\label{trgb_3.ps}
\end{figure}

\begin{figure}
\plotone{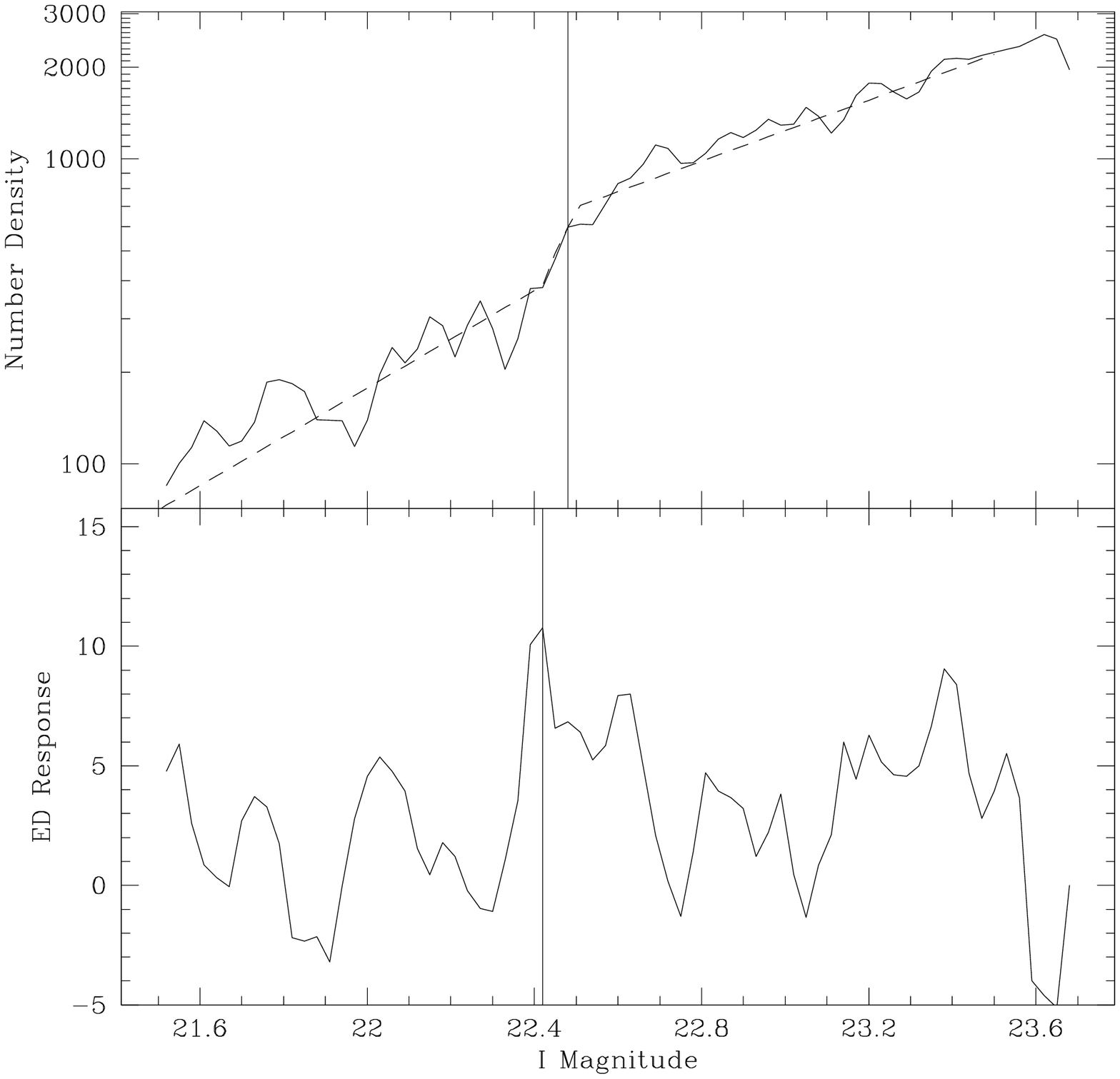}
\caption{Upper panel: Detection of the TRGB using ML method applied to Field 4. Lower panel: Detection of the TRGB using ED method applied to Field 4.}
\label{trgb_4.ps}
\end{figure}

\begin{figure}
\plotone{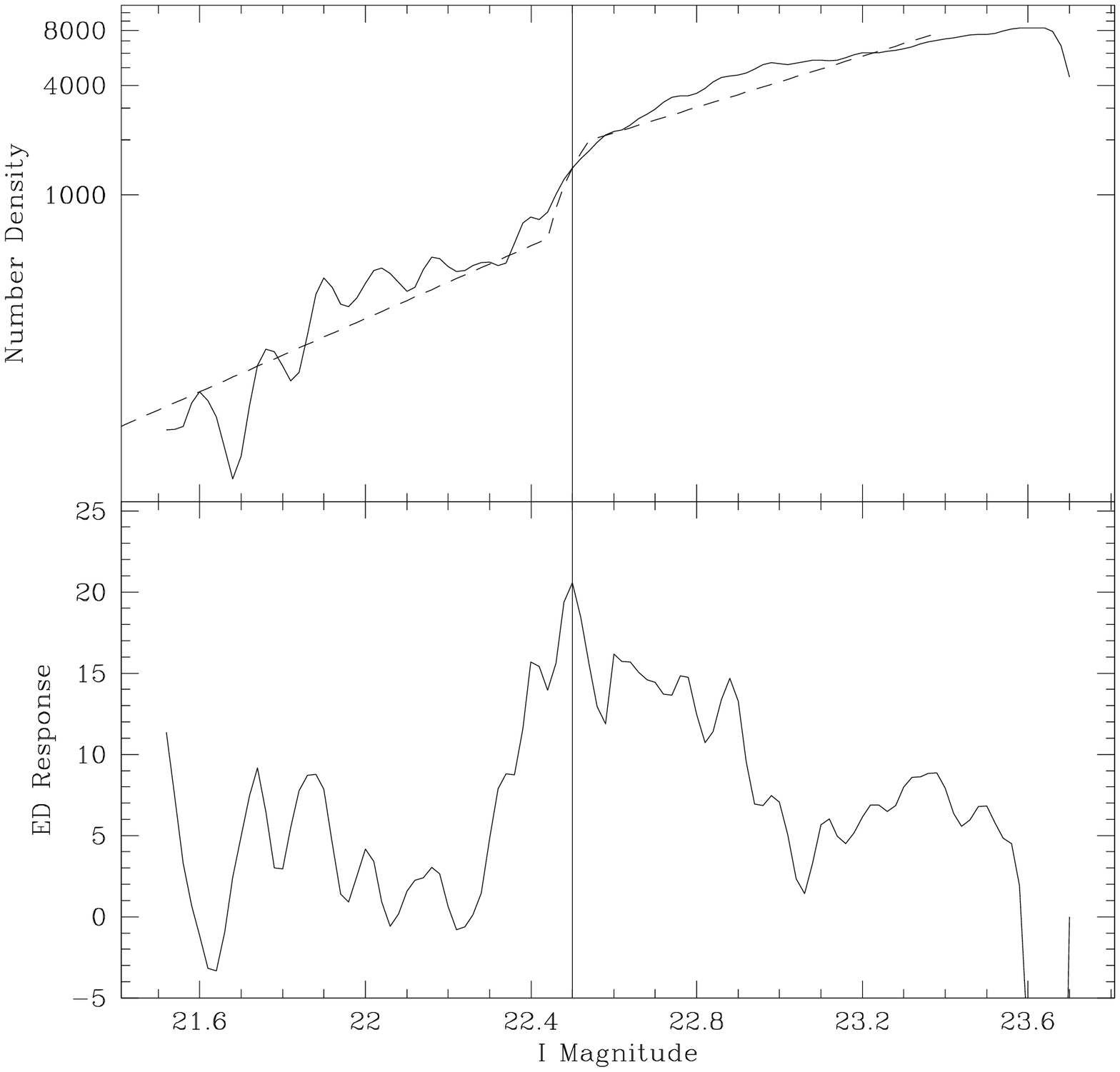}
\caption{Upper panel: Detection of the TRGB using ML method applied to Field 5. Lower panel: Detection of the TRGB using ED method applied to Field 5.}
\label{trgb_5.ps}
\end{figure}

\begin{figure}
\plotone{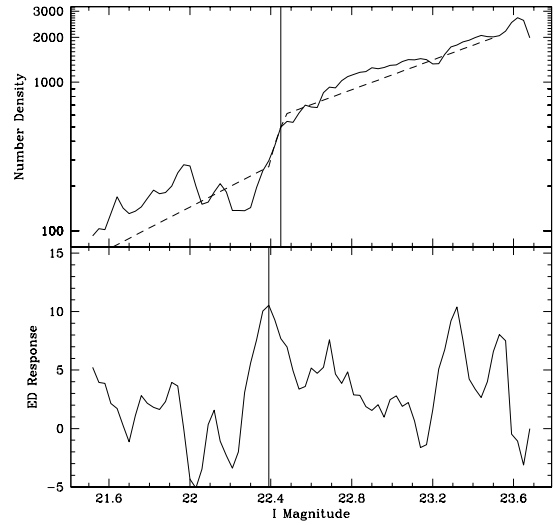}
\caption{Upper panel: Detection of the TRGB using ML method applied to Field 6. Lower panel: Detection of the TRGB using ED method applied to Field 6.}
\label{trgb_6.ps}
\end{figure}

The upper panels of Figures \ref{trgb_1.ps} - \ref{trgb_6.ps} show the detection of the RGB tip using the ML approach presented by \citet{2002AJ....124..213M}. The continuous line shows the observed RGB luminosity function, while the best fit is shown by a dashed line. The results of the detection are presented in columns 4 and 5 of Table \ref{tab1}. The lower panels of the same set of Figures show the detection of the RGB tip using the ED filter in a version similar to the one presented in \citet{1996ApJ...461..713S}. The continuous line shows the response of the ED filter, while the vertical line indicates the position of the center of the highest peak. The results of the measurements are reported in columns 2 and 3 of Table \ref{tab1}.

The discontinuity in the luminosity function due to the RGB tip is conspicuous in most cases, although a significant amount of contamination from AGB stars is affecting Fields 2 and 3, producing a rather smooth slope at the level of the RGB tip. The effect of an AGB contamination has been investigated in many studies, \citep[e.g., see][]{makarov:co,2004ApJ...606..869B}. The conclusion is that in most cases the RGB tip detection is quite insensitive to the effect of this contamination. This result is further confirmed by looking at the results presented here. Indeed, the RGB tip positions measured in Fields 2 and 3 do not significantly differ from the positions measured in any other field. 

To estimate the errors connected with the detection of the RGB tip, we adopted a bootstrap resampling strategy similar to the one presented in  \citet{2002AJ....124..213M}. The sample of stars chosen to represent the RGB was resampled 500 times, and the RGB tip measured for each realization. The r.m.s. of the results is then quoted in columns 3 and 5 of Table \ref{tab1}, for ED and ML methods, respectively.
\begin{table}
\begin{tabular}{c|cc|cc}
\tableline
\tableline
 & \multicolumn{2}{c|}{Edge detector} & \multicolumn{2}{c}{Maximum likelihood} \\
Field & I$_{RGBT}$  & $\sigma$ & I$_{RGBT}$ & $\sigma$ \\
\tableline
1 &  22.48 & 0.09 & 22.50 & 0.03 \\
2 &  22.40 & 0.03 & 22.48 & 0.02\\
3 &  22.48 & 0.06 &  22.50 & 0.02\\
4 &  22.42 & 0.16 &  22.48 & 0.06\\
5 &  22.50 & 0.10 &  22.50 & 0.02\\
6 &  22.39 & 0.12 &  22.45 & 0.08\\
\tableline
\end{tabular}
\caption{Results of the measurements of the magnitude of the RGB tip.\label{tab1}}

\end{table}

\subsection{Distance modulus}
\label{calib}

The first calibration of the absolute magnitude of the RGB tip dates back to the early 1990's.
\citet{1993ApJ...417..553L} defined the distance modulus based on the RGB tip as $$(m-M)_I=I_{TRGB}+BC_I-M_{bol,TRGB}$$ where $BC_I$ is the bolometric correction to the I magnitude, and $M_{bol,TRGB}$ is the bolometric magnitude of the TRGB. $BC_I$ and $M_{bol,TRGB}$ are given in \citet{1990AJ....100..162D} as $\rm{BC_I}=0.881-0.243(V-I)_0$
and $M_{bol}=-0.19\rm{[Fe/H]}-3.81$. These calibrations are based on the distance scale of \cite{1990ApJ...350..155L} where the magnitude of RR Lyrae stars is $M_V(RR)=0.82 + 0.17 \rm{[Fe/H]}$.
All these relations are based on a small sample of RGB stars observed in a few template globular clusters, and they only cover the range $-2.17 < \rm{[Fe/H]} < -0.71$. 
An extensive set of computer simulations was performed by \citet{1995AJ....109.1645M} to test for possible systematic effects on the detection of the RGB tip. The authors found that a reasonable lower limit to the number of stars within 1 magnitude from the tip is 50. Below this level, strong biases can affect the determined magnitude of the tip. Note that in the sample of \citet{1990AJ....100..162D} the number of stars within 1 magnitude from the tip is never larger than 20, and can be as low as 2.

A significant improvement on this situation was presented by \citet{2001ApJ...556..635B}. In their work, the authors derive a new calibration of the magnitude of the tip in the form $M_I^{TRGB}=0.14\rm{[Fe/H]}^2 + 0.48\rm{[Fe/H]} -3.66$.
The result is based on an extensive sample of stars observed in different bands including the near-IR and presented in 
\citet{1999AJ....118.1738F,2000AJ....119.1282F}. Although based on a larger sample of stars than the one presented in \citet{1993ApJ...417..553L}, this calibration still does not meet the completeness criteria established by  \citet{1995AJ....109.1645M}. In addition, both this calibration and the one by \citet{1993ApJ...417..553L} require a knowledge of the metallicity of the underlying population, either measured independently or deduced from the color of the RGB, iterating through measurements of the distance and the metallicity.

%We will adopt this calibration in this paper for two reasons: first, it is based on the largest photometric database available to date, and second, it covers the metallicity range $-2.2 < \rm{[Fe/H]} -0.2$.

%The metallicity of NGC 300 lies in the range $-0.5 < \rm{[Fe/H]} < -0.3$. Using [Fe/H]=-0.3, the absolute magnitude of the RGB tip would then be $M_I^{TRGB}=-3.79$. 

%%
%
 %and the corresponding distance moduli for the six fields are presented in Table \ref{tab2}. The results are obtained assuming $E(B-V)=0.096 \pm 0.006$ \cite{Gieren:2005ej}.

%begin{table}
%\begin{tabular}{c|cc|cc}
%\tableline
%\tableline
% & \multicolumn{2}{c|}{Edge detector} & \multicolumn{2}{c}{Maximum likelihood} \\
%Field & $(m-M)_0$  & STD & $(m-M)_0$ & STD \\
%\tableline
%1 &  26.10 & 0.00 & 26.12 & 0.00 \\
%%2 &  26.02 & 0.00 & 26.10 & 0.00\\
%3 &  26.10 & 0.00 &  26.12 & 0.00\\
%4 &  26.04 & 0.00 &  26.10 & 0.00\\
%5 &  26.12 & 0.00 &  26.12 & 0.00\\
%6 &  26.01 & 0.00 &  26.07 & 0.00\\
%\tableline
%\end{tabular}
%\caption{Results of the measurements of the distance modulus. \label{tab2}}
%\end{table}

\begin{figure}
\plotone{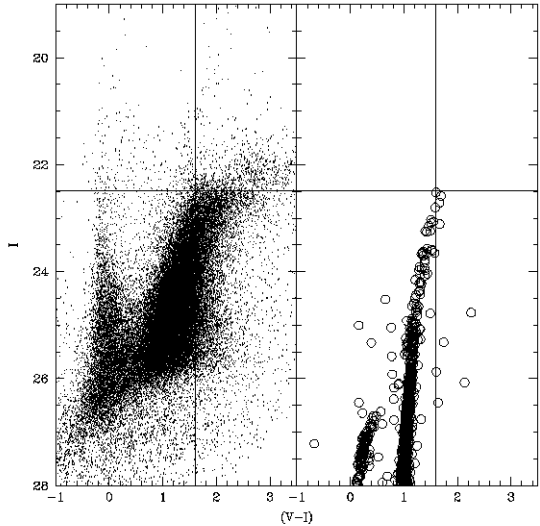}
\caption{ Left panel shows the CMD of NGC 300, right panel shows the CMD of $\omega$ Centauri. Vertical and horizontal lines indicate the color and the magnitude of the TRGB as measured in NGC 300.}
\label{omegacen.ps}
\end{figure}

The only calibration based on a sufficient number of stars is derived for $\omega$ Centauri by 
\citet{2001ApJ...556..635B}. According to this calibration, the absolute magnitude of the RGB tip is 
$M_I^{TRGB}=-4.04 \pm 0.12$ at a metallicity of ${\rm [Fe/H]} \sim -1.7$. This value is tied to the distance of the eclipsing binary OGLEGC 17 in $\omega$ Centauri \citep{2001AJ....121.3089T}, and it's completely independent from any other optical RR Lyrae distances. A possible source of uncertainty associated with this calibration is the wide and complex color/metallicity distribution observed in $\omega$ Centauri, but several studies have shown that the dominant population is rather metal-poor, and that the peak of the metallicity distribution is at ${\rm [Fe/H]} \sim -1.7$ \citep{2000ApJ...534L..83P,1996AJ....111.1913S}. In this work, we will adopt the value $M_I^{TRGB}=-4.04 \pm 0.12$. We note that this assumption is the reason behind our choice of the selection criteria we have adopted to define the RGB sample, as can be verified in Figure \ref{omegacen.ps}. The left panel of Figure \ref{omegacen.ps} shows the CMD of NGC 300, Field 2. Only 20 \% of the stars are plotted, for easier reading. The right panel shows the CMD of $\omega$ Centauri from \citet{2000A&AS..145..451R,2000A&AS..144....5R}. Horizontal and vertical lines show the position of the RGB tip as measured in NGC 300.  It is evident that it is possible to define in NGC 300 a sample of RGB stars that perfectly overlaps with the RGB of $\omega$ Centauri.

Assuming $E(B-V)=0.096 \pm 0.008$ \citep{2005ApJ...628..695G}, we derived distance moduli both with the ED and ML methods, and for the 6 ACS Fields. The results are presented in Table \ref{tab3}.
To estimate the errors attached to these measurements, we separate the errors connected to the detection of the tip and the photometric calibration ({\em internal} error) and the errors due to the extinction correction and the calibration of the absolute magnitude of the tip ({\em external} error). The errors due to the detection of the tip have already been discussed earlier in this Section. The errors connected with the conversion from ACS photometric system and BVI system can be quantified in 0.02 mag 
\citep{2005astro.ph..7614S}. The error attached to the $E(B-V)$ measurement provided by \citet{2005ApJ...628..695G} is 0.006 mag, which accounts for a total of 0.01 mag attached to $A_I$. Finally, the error in the absolute calibration is 0.12 mag \citep{2001ApJ...556..635B}, and it's basically determined by the uncertainty in the distance to $\omega$ Centauri \citep{2001AJ....121.3089T}. The total amount of {\em internal} errors attached to the different distance moduli computed for the six Fields are reported in columns 3 and 5 of Table \ref{tab3}, for ED and ML methods, respectively.

\begin{table}
\begin{tabular}{c|cc|cc}
\tableline
\tableline
 & \multicolumn{2}{c|}{Edge detector} & \multicolumn{2}{c}{Maximum likelihood} \\
Field & $(m-M)_0$  & $\sigma$ & $(m-M)_0$ & $\sigma$ \\
\tableline
1 &  26.35 & 0.09 & 26.37 & 0.03 \\
2 &  26.26 & 0.04 & 26.35 & 0.03\\
3 &  26.35 & 0.06 &  26.37 & 0.03\\
4 &  26.28 & 0.16 &  26.35 & 0.06\\
5 &  26.37 & 0.10 &  26.37 & 0.03\\
6 &  26.26 & 0.13 &  26.32 & 0.08\\
\tableline
\end{tabular}
\caption{Results of the measurements of the distance modulus. \label{tab3}}
\end{table}

To derive our final distance moduli, we computed a weighted mean of the measurements in the six Fields. The results are:
$$(m-M)_0=26.30 \pm 0.03 \pm 0.12 (ED)$$
and
$$(m-M)_0=26.36 \pm 0.02 \pm 0.12 (ML).$$

\section{Discussion}
\label{discussion}
\begin{figure}
\plotone{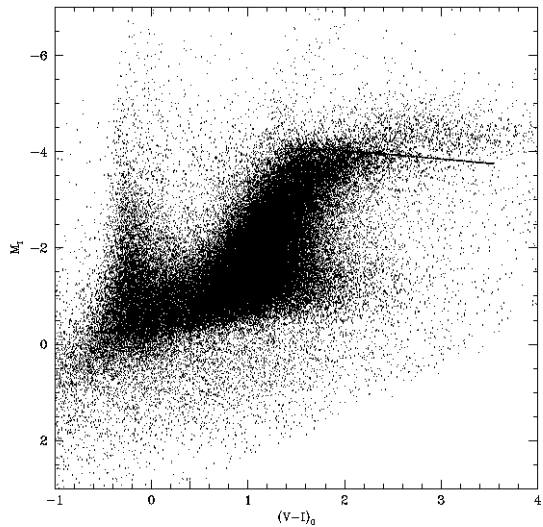}
\caption{CMD of NGC 300 in the absolute plane. The continuous line shows the color dependence of the TRGB according to \citet{2001ApJ...556..635B}}
\label{ferraro2.ps}
\end{figure}

Our selection of the sample of the stars representing the RGB is entirely motivated by our choice of the absolute calibration of the RGB tip. This approach actually limits the analysis to about 20 \%  of the total number of available RGB stars. 
As an alternative approach, one could choose to adopt a much larger sample of RGB stars, reaching the high-metallicity edge of the RGB. We argue that this approach would provide consistent results, but with a lower precision. This is shown in Figure \ref{ferraro2.ps}. In this Figure we plot the CMD of NGC 300, Field 2, in the absolute plane, using the distance and the reddening provided by \citet{2005ApJ...628..695G}. The continuous line shows the color dependence of the RGB tip according to \citet{2001ApJ...556..635B}. It is evident that the slope of the function $M_I^{TRGB} vs. (V-I)_0$ reproduces very closely the observed data. On the other hand, using the high-metallicity part of the CMD would introduce additional errors  due the still uncertain slope of the high-metallicity extension of the calibration.

Another issue that should be given attention to is the age of the underlying population used to define the RGB sample. Whenever the RGB tip technique is applied to a composite stellar population, the possibility of biases arises, due to the fact that the presence of a well-developed and populated RGB does not necessarily imply the presence of a globular cluster-like population, while the calibration of the absolute magnitude of the RGB tip relies completely on a sample of globular clusters. 
\citet{2004ApJ...606..869B} reported that the RGB distances are rather insensitive to the stellar populations provided most of the stars are more metal poor than $\rm{[Fe/H]}=-0.3$ and that there is not a strong star formation burst between 1 and 2 Gyr. \citet{2005MNRAS.357..669S} extended this analysis to real cases, and showed that applying the standard technique for RGB tip distances to the LMC and to the SMC could result in significant deviations from the real value, due to the underestimation of the correct metallicity.  We argue that the TRGB method can be safely applied to NGC 300, without introducing age- or metallicity-related biases. Indeed, \citet{2004AJ....127.1472B} have shown that the star formation history of this galaxy has been rather uniform throughout all its life, and they found no indication for an increased star formation rate at young ages, except for a possible final burst at 200-100 Myr.  Besides, both \citet{2004AJ....127.1472B} and the results of the Araucaria project \citep{2002ApJ...567..277B, 2003ApJ...584L..73U} show that the metallicity of NGC 300 has probably been lower than  ${\rm [Fe/H]}=-0.5$ for the whole life of the galaxy.

The result presented in this paper is fully consistent with the results recently derived by \citet{2005ApJ...628..695G}, based on the luminosity of Cepheids variable stars. Our distance modulus is also consistent with the one derived by \citet{2004AJ....127.1472B}, provided the difference in the adopted reddening correction is taken into account. Indeed, the observed magnitude of the RGB tip that we derived is consistent within the errors with the value $I_{TRGB}=22.52 \pm 0.02$ measured by \citet{2004AJ....127.1472B}, but the authors then apply a reddening correction $E(B-V)=0.013$ \citep{1998ApJ...500..525S}, which is much lower than the value adopted in this paper, resulting in a distance modulus $(m-M)_0=26.56 \pm 0.07 \pm 0.13$. Similar considerations apply to the results published by \citet{2005A&A...431..127T}, although in this case we do not know what is the adopted calibration of the absolute magnitude of the RGB tip, and the reddening correction applied. 
 
On the other hand, the results presented here show a significant discrepancy with the measurements of 
 \citet{2004ApJ...608...42S}, who published a distance modulus $(m-M)_0=26.65 \pm 0.09$. The total difference between this value and our value is $\sim 0.3$ magnitudes. Half of this difference can be explained by the different assumption of the reddening, as in the case of the distance presented by  
\citet{2004AJ....127.1472B}, but a further difference of $\sim 0.16$ remains to be explained. It appears that this difference can be accounted for by the difference in the estimated level of the RGB tip, measured at $I_{TRGB}=22.49 \pm 0.01$ in this paper, and at $I_{TRGB}=22.62 \pm 0.07$ by \citet{2004ApJ...608...42S}. It is difficult to provide an explanation for this difference, but a value of $I_{TRGB}=22.60$ is not compatible with our data. Besides, it is interesting to notice that the data analyzed by \citet{2004ApJ...608...42S} were also analyzed by \citet{2004AJ....127.1472B}, indicated as field F3.  Both groups determined the RGB tip around 22.6, but they also warned the reader that the field analyzed was poorly populated, and that the determination could be uncertain. Indeed, \citet{2004AJ....127.1472B} rejected the result derived from this WFPC2 field as non reliable. \citet{2004AJ....127.1472B} also analyzed an additional field, indicated as field F1, and for that field they derived the already quoted value of $I_{TRGB}=22.52 \pm 0.02$, in agreement with our determination. Our conclusion is that WFPC2 and ACS measurements agree within the errors when sufficient number of stars are used, as is the case of field F3 of  \citet{2004AJ....127.1472B}.

Finally, \citet{2004ApJ...608...42S} also reported that the Cepheids distance to NGC300, based on the measurements of \citet{1992ApJ...396...80F}, is $(m-M)_0=26.63 \pm 0.06$, but using the calibration of 
\citet{1999AcA....49..201U} the distance would be $(m-M)_0=26.53 \pm 0.05$, which would be in agreement with our determination if our value for the reddening would be used.

\section{Conclusions}
\label{conclusions}
We have presented a new measurement of the distance to NGC 300 based on the deepest available photometry catalog, obtained with the Advanced Camera for Survey on board the Hubble Space Telescope. We have used both edge-detection and maximum likelihood methods, and we have applied the methods independently to six different ACS Fields. All the Fields give consistent results. We have also discussed the possibility of biases in our results related to the application of the TRGB method to a composite stellar population, and we have concluded that NGC 300 is likely to be a case in which this distance estimator can be safely applied. Our result is fully consistent with the recent distance determination from near-infrared photometry of Cepheids variables \citep{2005ApJ...628..695G}. Since their result is tied to an assumed LMC distance modulus of 18.50, our independent TRGB distance determination of NGC 300 supports a distance of LMC of, or very close to, 18.50. The distance modulus we derive is also consistent with other recent determinations based on the TRGB \citep{2005A&A...431..127T,2004AJ....127.1472B} if our reddening value is used in these studies; however, our present determination has succeeded in reducing the internal errors of the result by a factor $\sim 3$.

\acknowledgements{WP and GP gratefully acknowledge support for this work from the Chilean FONDAP Center for Astrophysics 15010003 and the Polish KBN grant No 2P03D02123. Support for program \# GO-9492  was provided by NASA through a grant from the Space Telescope Science Institute, which is operated by the Association of Universities for Research in Astronomy, Inc., under NASA contract NAS 5-26555. We would like to thank the referee for useful suggestions and comments that helped improve this paper.}

\end{document}